\documentclass[amsmath,12pt,amssymb,preprint,nofootinbib]{revtex4}
\usepackage{amsfonts} %\usepackage{citesort}
\usepackage{graphicx} % Include figure files
\usepackage{epsfig}
\usepackage{multirow}
\usepackage{bm}% bold math

\begin{document}

\title{\textbf{QCD sum rule analysis of Bottomonium ground states}}
\author{Pallabi Parui}
\email{pallabiparui123@gmail.com} 
\author{Sourodeep De}
\email{sourodeepde2015@gmail.com}
\author{Ankit Kumar}
\email{ankitchahal17795@gmail.com} 
\author{Amruta Mishra}
\email{amruta@physics.iitd.ac.in}  
\affiliation{Department of Physics, 
Indian Institute of Technology, Delhi, New Delhi - 110016}

\begin{abstract}
 The in-medium masses of the bottomonium ground states [$1S$ ($\Upsilon (1S), 
\eta_b$) and $1P$ ($\chi_{b0},\chi_{b1}$)] are investigated in the strongly magnetized nuclear medium, using the QCD sum rule framework. 
In QCD sum rule approach, the mass modifications are calculated 
in terms of the medium modifications of the scalar and twist-2 
gluon condensates, which are obtained from the medium change
of a scalar dilaton field, $\chi$ within a chiral effective 
model. The gluon condensate is associated with the trace-anomaly 
of QCD, and it is incorporated into the chiral SU(3) model 
by the scalar dilaton field $\chi$ using 
a scale-invariance breaking logarithmic potential. 
The in-medium masses of the  bottomonium ground states
are observed to decrease with increasing density. 
P-wave states are observed to have more appreciable mass-shifts 
than the S-wave states. In the present investigation, the effects 
of spin-mixing between 1S bottomonium states, $\Upsilon(1S)$ 
and $\eta_b$ are taking into account in presence of an external 
magnetic field. The contribution of magnetic fields are seen 
to be dominant via spin-magnetic field interaction effects, 
which leads to an appreciable rise and drop in the in-medium 
masses of the longitudinal component of vector $1S$ state ($\Upsilon$) 
and pseudoscalar state ($\eta_b$) respectively. 
For zero magnetic field, the effects of baryon density 
on the bottomonium ground states in isospin asymmetric 
nuclear medium has been investigated which can have observable
consequences for the production of the open and hidden bottom
meson states. There is observed to be large contributions
to the masses of the longitudinal component of the
vector bottomonium state, $\Upsilon (1S)$ and pesudoscalar
state $\eta_b$, which might show in the dilepton spectra
in non-central ultra-relativistic heavy ion collision experiments
at RHIC and LHC, where the produced magnetic field is huge.
\end{abstract}
\maketitle
\vspace{-1cm}
\section{Introduction}
The study of the in-medium properties of hadrons is an important area of research in the physics of strongly interacting matter. The study of the heavy flavor hadrons \cite{Hosaka} has attracted a lot of attention due to its relevance in the ultra-relativistic heavy ion collision experiments. Recently, heavy quarkonia ($\overline{q}q; q= c,b$) under extreme conditions of matter, i.e., high density and/or high temperature, have been investigated extensively. The medium, created in the relativistic high energy collisions between heavy nuclei, affect the particles masses and decay widths, which have further observable impacts, e.g., particle production ratio etc. In the non-central heavy ion collision experiments, strong magnetic fields are expected to be produced \cite{kharzeev,fukushima,skokov,deng}. The magnetic fields produced, have been estimated to be huge in RHIC, BNL and in LHC, CERN \cite{tuchin}. However, the time evolution of the magnetic field produced in such experiments 
requires the detailed knowledge of the electrical conductivity 
of the medium and careful treatment of the solutions of magneto-hydrodynamic 
equations \cite{tuchin} and is still an open question. The study of the effects of strong magnetic fields on the in-medium properties of hadrons has 
initiated a new area of research in the heavy ion physics. \\
The heavy quarkonium (charmonium and bottomonium) states have been investigated in the literature using the potential models \cite{eichten1,eichten2,radfort}, the QCD sum rule approach \cite{klingl,kim,ko,am82,pallabi}, coupled channel approach \cite{molina}, the quark-meson coupling model \cite{krein,tsushima}, a chiral effective model \cite{amupsarx, am981, am90}, and a field theoretic model 
for composite hadrons \cite{am102, am95}.
In the present work, we study the masses of the S-wave (Vector,
$\Upsilon (1S)$, Pseudoscalar, $\eta_b$) and P-wave (scalar, $\chi_{b0}$, axial vector $\chi_{b1}$) bottomonium ground states, within the  magnetized asymmetric nuclear medium using the formalism of QCD sum rule. The S-wave and P-wave charmonium ground states in magnetized matter have already been studied \cite{cho91, pallabi} by the sum rule formalism. The medium effects are incorporated through the QCD gluon condensates \cite{schechter}, up to dimension 4, in terms of the medium modifications of scalar fields within a chiral SU(3) model based on the non-linear realization of chiral ${SU(3)}_L\times{SU(3)}_R$ symmetry.\\

 The open heavy flavor mesons, namely the open charm and the open bottom mesons, have also been studied within the QCD sum rule approach \cite{arata,wang, gubler}, the model for composite hadrons \cite{amarx} and the chiral model, without magnetic field \cite{am23, amd91}, and also with magnetic field \cite{am97, am98}. The in-medium masses of the light vector mesons have been studied using the QCD sum rule approach \cite{hatsuda}, where the medium modifications come through the non-strange and strange light quark condensates and the scalar gluon condensates, calculated within the chiral effective model, in strange asymmetric matter, without the effect of magnetic field \cite{am91}, and in nuclear medium with the effect of magnetic field \cite{am100}. In QCD sum rule approach, the mass modifications of the hidden heavy flavor mesons (charmonium and bottomonium) are found through the medium modifications of the scalar and the twist-2 gluon condensates calculated in a chiral effective model. The open heavy flavor mesons have their mass modifications in terms of both the light quark condensates (because of the light quark flavor present in their quark structure) as well as gluon condensates simulated within the chiral model. By finding out the mass modifications of the charmonium (bottomonium) and open charm (bottom) mesons within the medium, modifications of their decay widths have also been calculated using a field theoretic model of composite hadrons in presence of magnetic field \cite{am102, amsm1, amsm2} also using a light quark- anti-quark pair creation model, namely $^3P_0$ model \cite{amal}. These studies have important observable consequences in the relativistic heavy ion collision experiments, with the current focus on the heavy flavor meson in magnetized matter \cite{machado, matheus, suzuki}. There have also been a numbers of finite temperature studies of heavy quarkonia within the QCD sum rule framework. The properties of strange mesons have been investigated in magnetized matter, within the chiral SU(3) model \cite{anuj} and in the field theoretic model \cite{amsm3}.         
\\
In section II, the chiral SU(3) model has been described briefly; 
section III introduces with the Quantum-Chromodynamical (QCD) Sum Rule 
to find the in-medium masses of the lowest bottomonium states and the 
mass shifts due to spin-mixing effects; in section IV, the results 
of the in-medium masses and their shifts at various conditions 
are discussed; section V summarizes the findings of this work.

\section{The Chiral ${SU(3)}_L\times{SU(3)}_R $ Model}

In-medium masses of the bottomonium ground states are computed within the QCD sum rule approach, in terms of the gluon condensates. These condensates, in the present study, are calculated in a chiral effective SU(3) model \cite{papa59}. The chiral model is based on the non-linear realization of chiral symmetry \cite{weinberg, coleman, bardeen}, and the scale invariance breaking of QCD \cite{papa59, am69, zschi}. An effective Lagrangian, based on the non-linear realization of chiral symmetry has been employed here, with a logarithmic potential in scalar dilatation field \cite{erik}, $\chi$ mimics the scale-invariance breaking of QCD. The chiral $ {SU(3)}_L\times{SU(3)}_R$ model Lagrangian density has the following general form \cite{papa59},
\begin{equation}
   \mathcal{L}=\mathcal{L}_{kin}+\mathcal{L}_{BM}+\mathcal{L}_{vec}+\mathcal{L}_0+\mathcal{L}_{scale-break}+\mathcal{L}_{SB}+\mathcal{L}_{mag} 
\end{equation}
in the above expression, $\mathcal{L}_{kin}$ is the kinetic energy of the baryons and the mesons; $\mathcal{L}_{BM}$ represents the baryon-mesons (spin-0 and spin-1) interactions; $ \mathcal{L}_{vec}$ , contains the quartic self-interactions of the vector mesons and their couplings with the scalar ones; $\mathcal{L}_0$ incorporates the spontaneous chiral symmetry breaking effects via meson-meson interactions; $\mathcal{L}_{scale-break}$ is the scale symmetry breaking logarithmic potential;  the explicit symmetry breaking term, $\mathcal{L}_{SB}$; finally the magnetic field effects on the charged and neutral baryons in the nuclear medium are given by \cite{amupsarx, am981, am97, am98, broderik, prakash, wei, guang},
\begin{equation}
\mathcal{L}_{mag}=-\frac{1}{4}F_{\mu\nu}F^{\mu\nu}-q_i{\bar{\psi}}_i\gamma_\mu A^\mu\psi_i-\frac{1}{4}\kappa_i\mu_N{\bar{\psi}}_i\sigma^{\mu\nu}F_{\mu\nu}\psi_i
\end{equation}
where,  $\psi_i$ is the baryon field operator $( i = p, n)$, in the case of nuclear matter, the parameter, $\kappa_i$ here is related to the anomalous magnetic moment of the i-th baryon ($p, n$) (\cite{broderik} - \cite{paoli}), with $\kappa_p = 3.5856$ and $\kappa_n = -3.8263$, are the gyromagnetic ratio corresponding to the anomalous magnetic moments (AMM) of the proton and the neutron respectively. Thus, in the magnetized nuclear medium, magnetic field has contributed through the Landau energy levels of the charged particles \cite{ivanov}, and through the non-zero anomalous magnetic moments of the nucleons \cite{ivanov, paoli}. 
\\
In the chiral model, mean-field approximation is adopted, where the meson fields are treated as classical fields. The expectation values of the fields within the system, have non-zero contribution only for the vector (time-component) and scalar meson fields and zero for the other meson (pseudoscalar, axial vector) fields \cite{papa59}.\\

The scalar dilatation field, $\chi$  simulates the scalar gluon condensate $ \langle \frac{\alpha_s}{\pi} G_{\mu\nu}^a$ $G^{a\mu\nu}\rangle$, as well as the twist-2 gluon operator $ \langle \frac{\alpha_s}{\pi} G_{\mu\sigma}^a$ $G_{\nu}^{a\space \sigma}\rangle$, within the model. The energy momentum tensor, $T_{\mu\nu}$ derived from the $\chi$-terms in the chiral model Lagrangian density \cite{am82} thus obtained,
\begin{equation}
 T_{\mu\nu}=\left(\partial_{\mu}\chi\right)\left(\frac{\partial\mathcal{L}}{\partial\left(\partial^\nu\chi\right)}\right)- g_{\mu\nu}\mathcal{L}_\chi
\end{equation}      
The QCD energy momentum tensor, in the limit of current quark masses, contains a symmetric trace-less part and a trace part, as given below \cite{morita, cohen}, 
\begin{equation}
T_{\mu\nu}=-ST\left(G_{\mu\sigma}^aG_\nu^{a\sigma}\right)+\frac{g_{\mu\nu}}{4}\left(\sum_{i}m_i\overline{q}_i q_i+\langle \frac{\beta_{QCD}}{2g}G_{\sigma k}^a G^{a\sigma}_k\rangle\right) 
\end{equation}
with the leading order QCD $\beta$ function \cite{am82}, $\beta_{QCD}(g) = -\frac{g^3}{(4\pi)^2} (11-\frac{2}{3} N_f )$, by taking the 3 color quantum numbers of QCD, and no. of flavors, $N_f=3$. Here, $m_i$'s $(i= u, d, s)$ are the current quark masses.\\ Writing the medium expectation value of the twist-2 gluon operator as, 
\begin{equation}
     \langle \frac{\alpha_s}{\pi} G_{\mu\sigma}^a G_{\nu}^{a\space \sigma}\rangle = \left(u_\mu u_\nu-\frac{g_{\mu\nu}}{4}\right)G_2
\end{equation}
where $u_\mu$ is the 4-velocity of the nuclear medium, taken to be at rest \cite{am82, pallabi} in the present investigation, $u_\mu=\left(1,\ 0,\ 0,\ 0\right)$; the QCD energy momentum tensor then reads 
\begin{equation}
    T_{\mu\nu}=-\frac{\pi}{\alpha_s}\left(u_\mu u_\nu-\frac{g_{\mu\nu}}{4}\right)G_2+\frac{g_{\mu\nu}}{4}\left(\sum_{i}m_i\overline{q}_i q_i+\langle \frac{\beta_{QCD}}{2g}G_{\sigma k}^a G^{a\sigma}_k\rangle\right)
\end{equation}
Comparing the expressions of energy momentum tensor in equation(6) and in equation(3), one obtains the expressions for $G_2$ (the twist-2 component) and the scalar gluon condensate by multiplying both sides with $\left(u_\mu u_\nu-\frac{g_{\mu\nu}}{4}\right)$ and $g^{\mu\nu}$ respectively. These are given by-

\begin{equation}
G_2 = \frac{\alpha_s}{\pi}\Bigg[-(1-d+4k_4)(\chi^4-\chi_0^4)-
\chi^4 \ln\left(\frac{\chi^4}{\chi_0^4}\right)+\frac{4}{3}d\chi^4\ln\left(\left(\frac{(\sigma^2-\delta^2)\zeta}{\sigma_0^2\zeta_0}\right)\left(\frac{\chi}{\chi_0}\right)^3\right)\Bigg]
\end{equation}
and, 
\begin{equation}
 \langle \frac{\alpha_s}{\pi} G_{\mu\nu}^a G^{a\mu\nu}\rangle
=\frac{8}{9}\Bigg[(1-d)\chi^4+\left(\frac{\chi}{\chi_0}\right)^2\left(m_\pi^2 f_\pi \sigma +\left(\sqrt{2}m_k^2f_k-\frac{1}{\sqrt{2}}m_\pi^2 f_\pi\right)\zeta\right)\Bigg]
\end{equation}
Thus, the expectation values of the scalar and the twist-2 gluon condensates depend on the in-medium values of the non-strange scalar field, $\sigma$, the strange scalar field, $\zeta$, the scalar-isovector field, $\delta$ [ if the current quark masses are taken to be non-zero ] besides the scalar dilaton field, $\chi$, within the chiral $ SU(3) $ model. By deriving the Euler Lagrange's equations of motion from the effective model Lagrangian under mean-field approximation, the coupled equations of motion in these fields are obtained, incorporating the effects of density, isospin asymmetry, anomalous magnetic moments, and magnetic field of the nuclear medium under study.  
The coupled equations of motion in $\sigma, \zeta, \delta,$ and $\chi $ are,
\begin{eqnarray}
   && k_0\chi^2\sigma-4k_1(\sigma^2+\zeta^2+\delta^2) \sigma 
-2k_2(\sigma^3+3\sigma\delta^2)-2k_3 \chi\sigma\zeta
\nonumber \\ &-&\frac{d}{3}\chi^4\left(\frac{2\sigma}{\sigma^2
-\delta^2}\right)+\left(\frac{\chi}{\chi_0}\right)^2 m_\pi^2 f_\pi 
- \sum g_{\sigma i} \rho_i^s = 0
\end{eqnarray}
\begin{eqnarray}
&&     k_0\chi^2\zeta -4k_1(\sigma^2+\zeta^2+\delta^2)\zeta-4k_2\zeta^3-k_3 \chi(\sigma^2-\delta^2)-\frac{d}{3}\frac{\chi^4}{\zeta} \nonumber \\ 
&+&\left(\frac{\chi}{\chi_0}\right)^2\left(\sqrt{2}m_k^2f_k-\frac{1}{\sqrt{2}}m_\pi^2 f_\pi\right)-\sum g_{\zeta i} \rho_i^s = 0 
\end{eqnarray}
\begin{eqnarray}
  &&  k_0\chi^2\delta -4k_1(\sigma^2+\zeta^2+\delta^2)\delta - 2k_2(\delta^3+3\sigma^2\delta) +k_3\chi\delta\zeta \nonumber \\ &+&\frac{2}{3}d\chi^4\left(\frac{\delta}{\sigma^2-\delta^2}\right)-\sum g_{\delta i}\rho_i^s =0
\end{eqnarray}
\begin{eqnarray}
 &&   k_0\chi(\sigma^2+\zeta^2+\delta^2)-k_3(\sigma^2-\delta^2)\zeta+ \chi^3 \left[1+4\ln\left(\frac{\chi^4}{\chi_0^4}\right)\right]+ (4k_4 - d) \chi^3 
\nonumber \\ &-&\frac{4}{3}d\chi^3 \ln\left[\left(\frac{(\sigma^2-\delta^2)\zeta}{\sigma_0^2\zeta_0}\right)\left(\frac{\chi}{\chi_0}\right)^3\right]+2\frac{\chi}{\chi_0^2}\left[m_\pi^2 f_\pi \sigma +\left(\sqrt{2}m_k^2f_k-\frac{1}{\sqrt{2}}m_\pi^2 f_\pi\right)\zeta\right] = 0.
\end{eqnarray}
in these equations, $\rho_i ^s$ is the scalar density for the  $i$-th  
baryon in the magnetized matter \cite{amupsarx}.   

\section{In-Medium Masses Within The QCD Sum Rule Approach}
In this section, the in-medium masses of the lowest-lying bottomonium S-waves 1S [$\Upsilon (1S)$, $\eta_b$  and P-waves, 1P [$\chi_{b0}$ and $\chi_{b1}$], are calculated within the QCD Sum Rule approach. In the QCD Sum Rule framework, the masses of these bottomonium ground states are obtained by using the medium modified scalar and twist-2 gluon condensates. The condensates are then calculated within the chiral SU(3) model in terms of the scalar fields modifications within the magnetized, iso-spin asymmetric nuclear matter.  
The in-medium mass squared, $m_i^{*2}$  for the i-type bottomonium ground state [ i= vector, pseudoscalar, scalar, and axial-vector ], in the QCD sum rule can be written as \cite{reinders}, 
\begin{equation}
    m_i^{*2}\simeq \frac{M_{n-1}^i (\xi)}{M_{n}^i (\xi)}-4m_b^2\xi
\end{equation}
    where $M_{n}^i $ is the $n$-th moment of the i-type meson, and, 
$\xi$ is the renormalization scale.
    Using the operator product expansion technique [OPE], the moment can be written as \cite{klingl, reinders}, 
    \begin{equation}
      M_{n}^i(\xi)=A_n^i(\xi)\left[1+ a_n^i(\xi)\alpha_s + b_n^i(\xi)\phi_b+c_n^i(\xi)\phi_c\right] 
    \end{equation}
Here, $A_n^i, a_n^i, b_n^i,$ and $c_n^i$ are the Wilson coefficients. The $A_n^i$ coefficients result from the bare-loop diagram of perturbative QCD, $ a_n^i$ are the contributions from the perturbative radiative corrections, and the coefficients, $b_n^i$ are related to the scalar gluon condensate through 
\begin{equation}
    \phi_b=\frac{4\pi^2}{9}\frac{\langle \frac{\alpha_s}{\pi} G_{\mu\nu}^a G^{a\mu\nu}\rangle}{\left(4m_b^2\right)^2}.
\end{equation}
By the replacement of the value of scalar gluon condensate, above equation can be written in terms of the scalar fields as,
\begin{equation}
    \phi_b=\frac{32\pi^2}{81\left(4m_b^2\right)^2}\Bigg[(1-d)\chi^4+ \left(\frac{\chi}{\chi_0}\right)^2\left(m_\pi^2 f_\pi \sigma +\left(\sqrt{2}m_k^2f_k-\frac{1}{\sqrt{2}}m_\pi^2 f_\pi\right)\zeta\right)\Bigg]
\end{equation}
Finally, the $c_n^i$ coefficients are associated with the twist-2 gluon condensates through 
\begin{equation}
    \phi_c=\frac{4\pi^2}{3\left(4m_b^2\right)^2 }G_2
\end{equation}
In terms of the scalar fields, expression for $\phi_c$ is,
\begin{eqnarray}
    \phi_c &= & \frac{4\pi \alpha_s}{3\left(4m_b^2\right)^2} 
\Bigg[-(1-d+4k_4)(\chi^4-\chi_0^4)-
\chi^4 \ln\left(\frac{\chi^4}{\chi_0^4}\right) \nonumber \\ 
&+&\frac{4}{3}d\chi^4\ln\left(\left(\frac{(\sigma^2-\delta^2)\zeta}{\sigma_0^2\zeta_0}\right)\left(\frac{\chi}{\chi_0}\right)^3\right)\Bigg]
\end{eqnarray}
The $\xi$-dependent parameters $m_b$ and $\alpha_s$, are the running bottom quark mass and the running coupling constant respectively, given below \cite{am82, reinders},
\begin{equation}
    \frac{m_b(\xi)}{m_b} = 1-\frac{\alpha_s}{\pi}\left[\frac{2+\xi}{1+\xi} \ln(2+\xi)-2\ln2\right]
\end{equation}
with $m_b\equiv m_b (p^2=-m_b^2)=4.23 $ GeV \cite{reinders85}, and 
\begin{equation}
    \alpha_s \left(Q_0^2+4m_b^2\right) = \alpha_s(4m_b^2) \Bigg/ \left(1+\frac{(33-2n_f)}{12\pi}\alpha_s(4m_b^2)\ln\frac{Q_0^2+4m_b^2}{4m_b^2}\right)
\end{equation}
where, $n_f=5$, $\alpha_s \left(4m_b^2\right) \simeq 0.15$ \cite{reinders85}, and $Q_0^2=4m_b^2 \xi$. \\
The Wilson coefficients $A_n^i, a_n^i, b_n^i,$ are given in ref.\cite{reinders} for different quantum numbers, $J^{PC}$ of particle states, for e.g., the scalar, vector, pseudoscalar, axial-vector channels, etc. The $c_n^i$'s are listed for the vector and pseudoscalar channels in \cite{klingl}, in case of S-wave charmonium ground states, and, for the P-waves (scalar and axial-vector), 
$c_n^i$'s are calculated using a background field technique 
in Ref. \cite{song}.
In the present work, in the presence of an external magnetic field,
we have considered the effects of spin-magnetic field interaction on 
the bottomonium 1S triplet and singlet states. The effects of spin-magnetic field interaction have been studied for the 1S charmonium states, vector $J/\Psi$ and pseudoscalar $\eta_c$ at finite magnetic fields \cite{pallabi, cho91, cho14, alford, suzuki17}. This leads to a mixing between $J/\Psi$ (1S) and $\eta_c$ states. At non-zero magnetic fields, the spin-magnetic field coupling leads to a mixing between the longitudinal component of the spin 1, $\Upsilon (1S)$ state and the spin 0, $\eta_b$ state. The masses of the longitudinal vector, $\Upsilon^{||}(1S)$  (pseudoscalar, $\eta_b$) bottomonium states, are seen to have a rise (drop) with increasing magnetic fields, when the spin-mixing effects are incorporated at finite magnetic fields. The effective masses of the $\Upsilon^{||}(1S)$ ( 1S triplet) and $\eta_b$ (1S singlet) by considering the shifts due to spin-magnetic field interaction, thus given by \cite{alford},
\begin{equation}
    m^{eff}_{\Upsilon (1S)}= m^*_{\Upsilon(1S)} + \Delta m_{sB},\;\;\;\; 
m^{eff}_{\eta_b}= m^*_{\eta_b} + \Delta m_{sB} 
\end{equation}
In the above equation, $m^*_{\Upsilon(1S)/\eta_b}$ denotes the in-medium masses of the S-waves bottomonium ground states calculated within QCD sum rule framework [equation. (13)], and $\Delta m_{sB}$ is the shift due to the spin-magnetic field interaction. Expression for the latter is given as below,
\begin{equation}
    \Delta m_{sB} = \frac{\Delta M}{2}\left((1+{\chi_{sB}}^2)^{1/2}-1\right),\\ \chi_{sB} = \frac{2g\mu_b B}{\Delta M}
\end{equation}
where, $\mu_b = (\frac{1}{3}e)/(2m_b)$ is the bottom quark Bohr magneton with the constituent bottom quark mass, $m_b = 4.7 $ GeV in the present work, $\Delta M = m^*_{\Upsilon} - m^*_{\eta_b}$, and g is chosen to be 2 (ignoring the effects of the anomalous magnetic moments of the bottom quark (anti-quark)). 
\section{Results and Discussions}
In the present work, in-medium masses of the lowest lying bottomonia are investigated within the QCD Sum rule (QCDSR) approach, which has extensively been 
applied to study the lowest lying hadronic resonances. The masses of the S-wave bottomonium ground states [vector, $\Upsilon (1^3S_1)$, pseudoscalar, $\eta_b (1^1S_0)$] and the P-wave bottomonium ground states [scalar $\chi_{b0} (1^3P_0)$, axial-vector $\chi_{b1} (1^3P_1)$] are investigated in the isospin asymmetric as well as symmetric nuclear matter, both in the presence of strong magnetic fields and in the absence of any magnetic field. The anomalous magnetic moments (AMM) of the nucleons (protons and neutrons) are considered while investigating the effects of magnetic fields, in the present work. In QCDSR, the masses are obtained by taking the ratio of two consecutive moments in the appropriate n-region, such that the stability can be found. \\In the sum rule calculations of the heavy quarkonia ($\overline{q}q, q=c, b$) masses, the moments [$M_n(\xi)$], in the QCD operator product expansion (up to dimension-4 here), are depended on the scalar gluon condensate ($\phi_b$) and the twist-2 gluon operator ($\phi_c$). The medium effects on the masses are incorporated through these condensates, which are further investigated within an effective chiral SU(3) model.\\ In the chiral model, the gluon condensates are simulated through the scalar dilaton field, $\chi$ which in turn mimics the scale symmetry breaking of QCD through a logarithmic potential. At non-zero magnetic fields, proton has Landau energy levels contribution, which are taken into consideration in the solution of scalar fields. For any given values of density ($\rho_B$), isospin asymmetry ($\eta$), and magnetic field (eB), coupled equations of motions in the scalar fields, $\sigma, \zeta, \delta,$ and $\chi$ are solved under the mean-field approximation of the chiral effective model Lagrangian. Then, the scalar and twist-2 gluon condensates are obtained in terms of these scalar fields from equations (8) and (7) respectively. Thereafter, using equations (15) and (16), $\phi_b$ and $\phi_c$ can be calculated. Thus the medium modifications of the scalar iso-scalar (strange $\zeta$, non-strange $\sigma$), scalar-isovector, $\delta$, scalar dilaton $\chi$, fields are projected in the final mass calculation of $\overline{b}b$ states through the condensates. The Wilson coefficients in $M_n(\xi)$ are obtained using QCD perturbation theory.\\In the present investigation, the value of $\xi$ is taken as 1 for the S-wave and 2.5 for the P-wave states, leading to the value of the running coupling constant, $\alpha_s$, to be 0.1411 [S states] and  0.1346 [P states], and of the running bottom quark mass, $m_b$ to be 4180.3 MeV and 4130.8 MeV, respectively.\\
The isospin asymmetry parameter is denoted by $\eta\left(=\frac{\rho_n-\rho_p}{2\rho_B}\right)$ [$\rho_n$ and $\rho_p$ : number density of the neutron and the proton respectively], and the nuclear matter saturation density, $\rho_0$ is 0.15 $fm^{-3}$ in the present work. Calculations are done in both the isospin symmetric ($\eta=0$), and asymmetric ($\eta=0.5$) nuclear matter at the baryonic density of, $\rho_B$ = $\rho_0$, 2$\rho_0$ and 4$\rho_0$, for finite values of the AMM of p and n.
Masses are calculated in the magnetized nuclear matter for the values of magnetic field, $|eB|$ = $4m_\pi^2$ and $12m_\pi^2$. Energy levels of protons have Landau levels contribution due to their electric charge, and both protons and neutrons have contribution due to their AMM in the magnetized nuclear matter. \\ The effects of asymmetric high density nuclear medium are also investigated in the present work, without considering the effects of magnetic field. The masses are calculated within the sum rule framework, by calculating the condensates in a chiral SU(3) model, without considering the magnetic field. Vacuum masses (at $\rho_B=0$) for all four states thus obtained, are shown in table \ref{table:1},
\begin{table}[h!]
\centering
\begin{tabular}{|c|c|}
\hline \textbf{Particle state} & \textbf{Vacuum mass (MeV)} \\[0.5ex] \hline
$ \eta_b (^1S_0) $ & 9681.479 \\ \hline
$\Upsilon (^3S_1)$ & 9751.249  \\ \hline
$\chi_{b0}(^3P_0)$ & 10573.429 \\ \hline
$\chi_{b1}(^3P_1)$ & 10812.121 \\ \hline
\end{tabular}
\caption{Vacuum Masses (MeV) of the S and P waves bottomonium ground states}
\label{table:1}
\end{table}

\begin{figure}
    \includegraphics[width=1.1\textwidth]{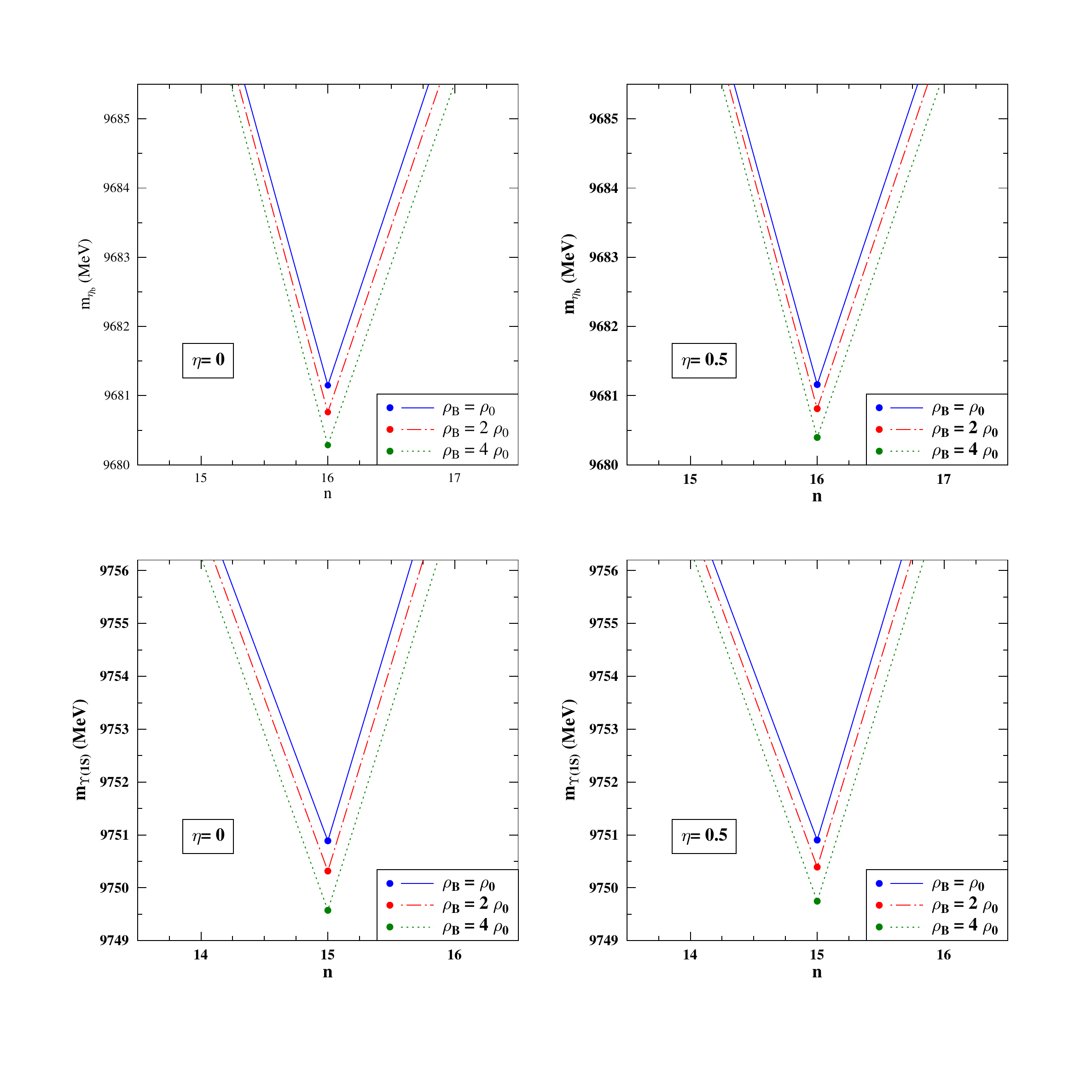}\hfill
\vskip -0.8in
    \caption{Masses (MeV) are plotted with variation in n, for 1S states, $\Upsilon$ and $\eta_b$ in the absence of magnetic field (eB = 0), at $\rho_B = \rho_0$, $2\rho_0$ and $4\rho_0$. Masses are plotted both at symmetric ($\eta$ = 0) and asymmetric ($\eta$ = 0.5) nuclear matter.}
    \label{fig:2a}
    \end{figure}
    \begin{figure}
    \includegraphics[width=1.1\textwidth]{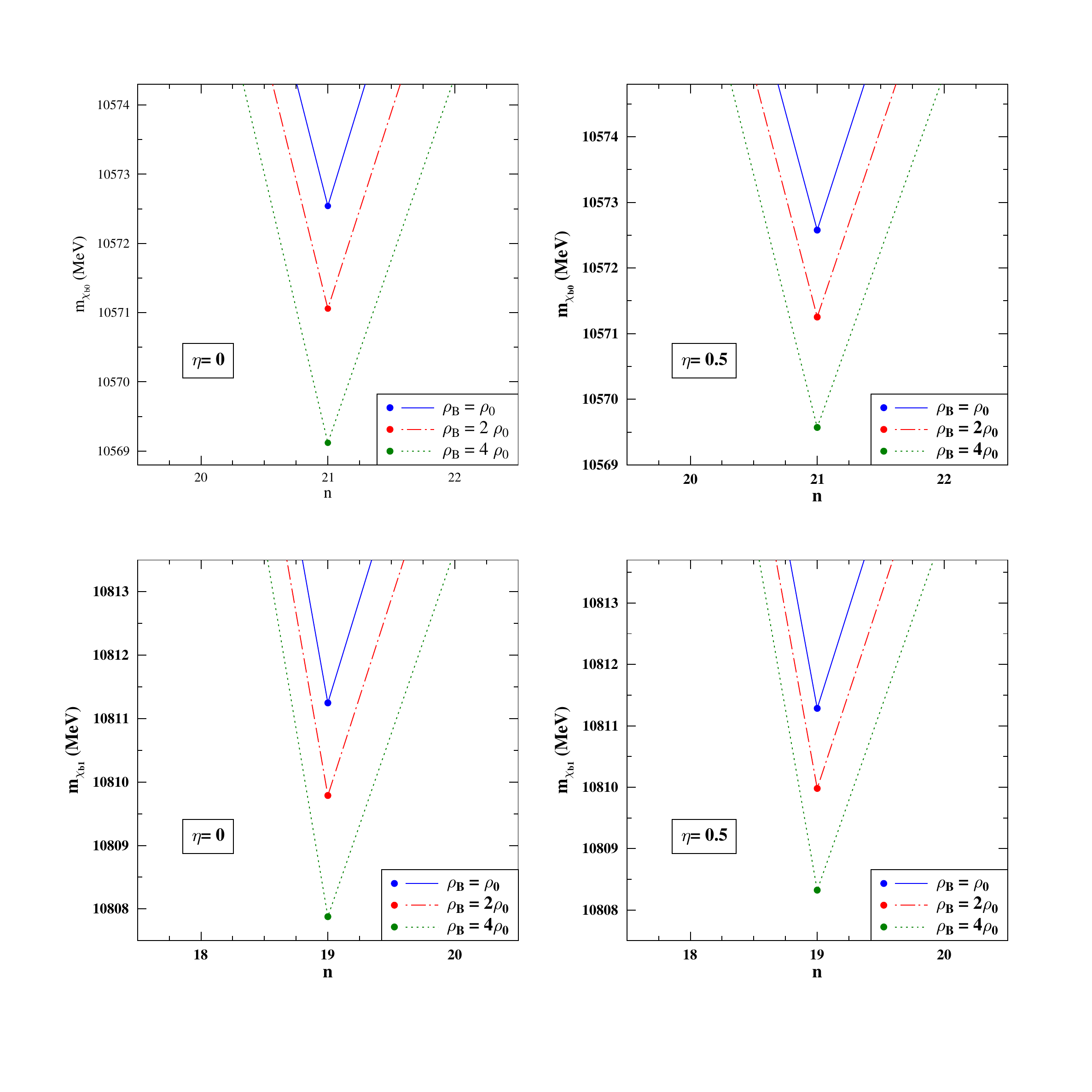}\hfill
\vskip -0.8in
    \caption{Masses (MeV) are plotted with variation in n, for 1P states, $\chi_{b0}$ and $\chi_{b1}$ in the absence of magnetic field, at $\rho_B = \rho_0$, $2\rho_0$ and $4\rho_0$. Masses are plotted both at symmetric ($\eta$ = 0) and asymmetric ($\eta$ = 0.5) nuclear matter.}
    \label{fig:2b}
\end{figure}
\begin{figure}
    \includegraphics[width=1.1\textwidth]{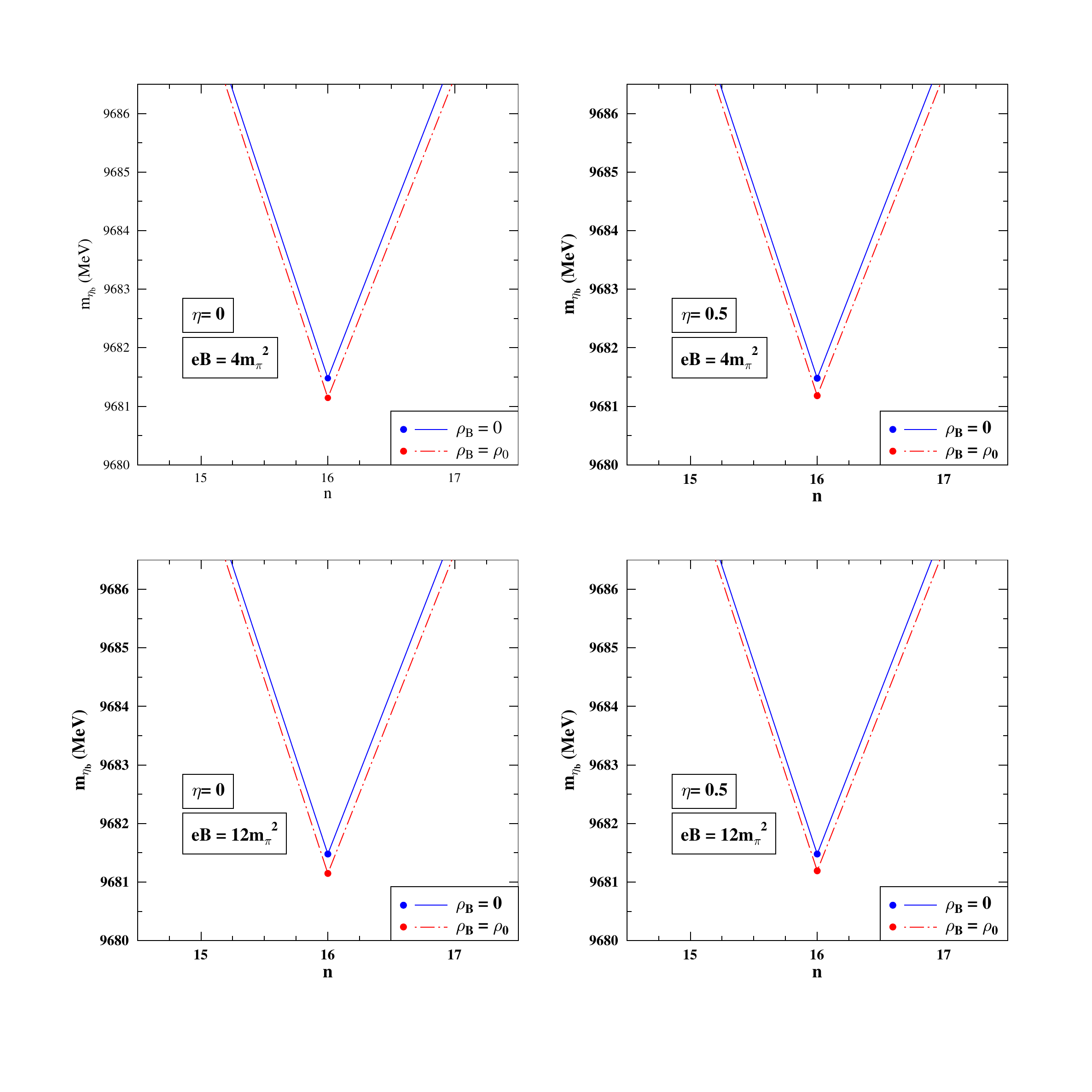}\hfill
\vskip -0.8in
    \caption{Masses (MeV) are plotted with variation in n, for the pseudoscalar 1S state, $\eta_b$ at non-zero magnetic field (eB = 4$m_{\pi}^2$ and 12$m_{\pi}^2$), for $\rho_B = 0$ and $\rho_0$. Masses are plotted both at symmetric ($\eta$ = 0) and asymmetric ($\eta$ = 0.5) nuclear matter.}
    \label{fig:3a}
    \end{figure}
    \begin{figure}
    \includegraphics[width=1.1\textwidth]{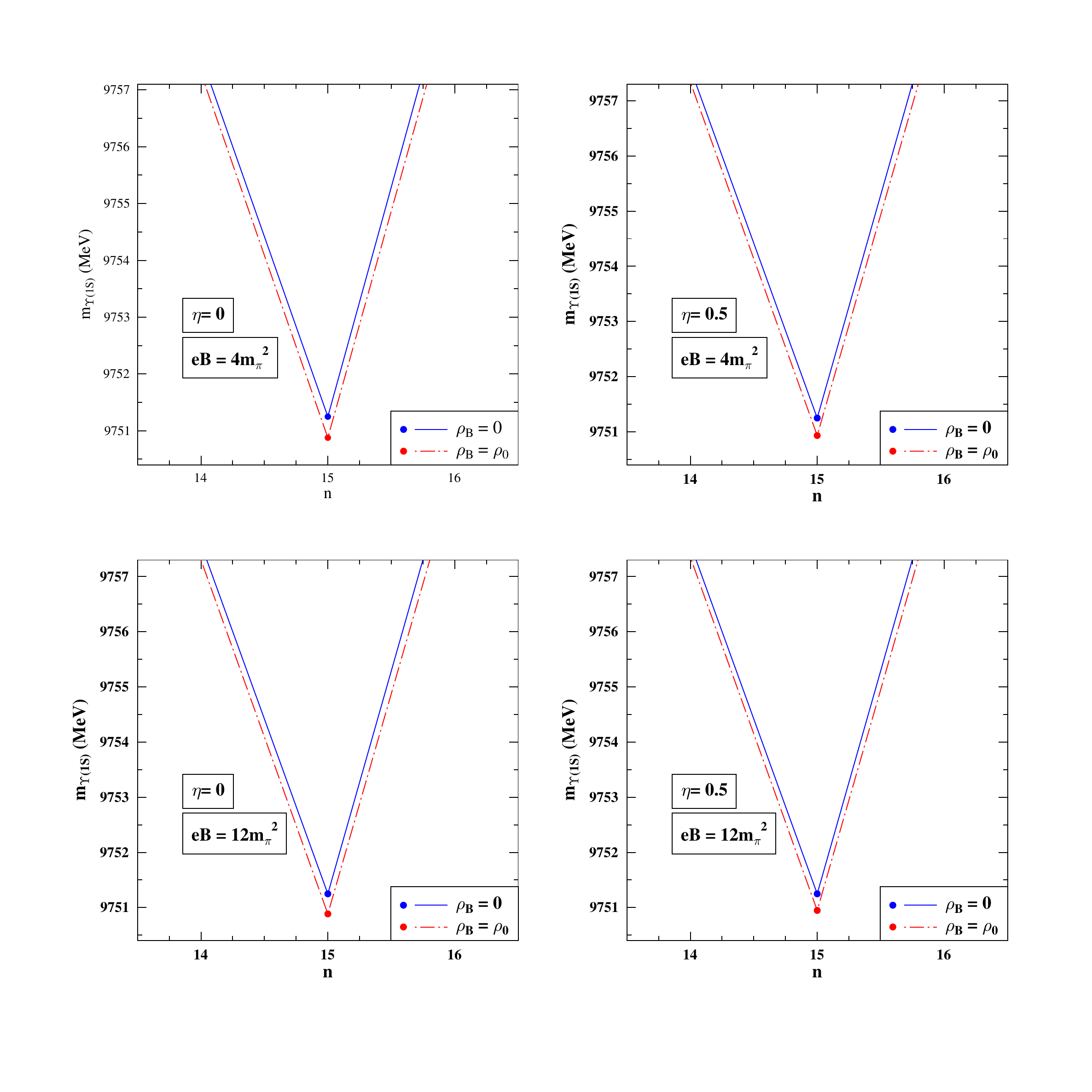}\hfill
\vskip -0.8in
    \caption{Masses (MeV) are plotted with variation in n, for the vector 1S state, $\Upsilon (1S)$ at non-zero magnetic field (eB = 4$m_{\pi}^2$ and 12$m_{\pi}^2$), for $\rho_B = 0$ and $\rho_0$. Masses are plotted both at symmetric ($\eta$ = 0) and asymmetric ($\eta$ = 0.5) nuclear matter.}
    \label{fig:3b}
\end{figure}

\begin{figure}
    \includegraphics[width=1.1\textwidth]{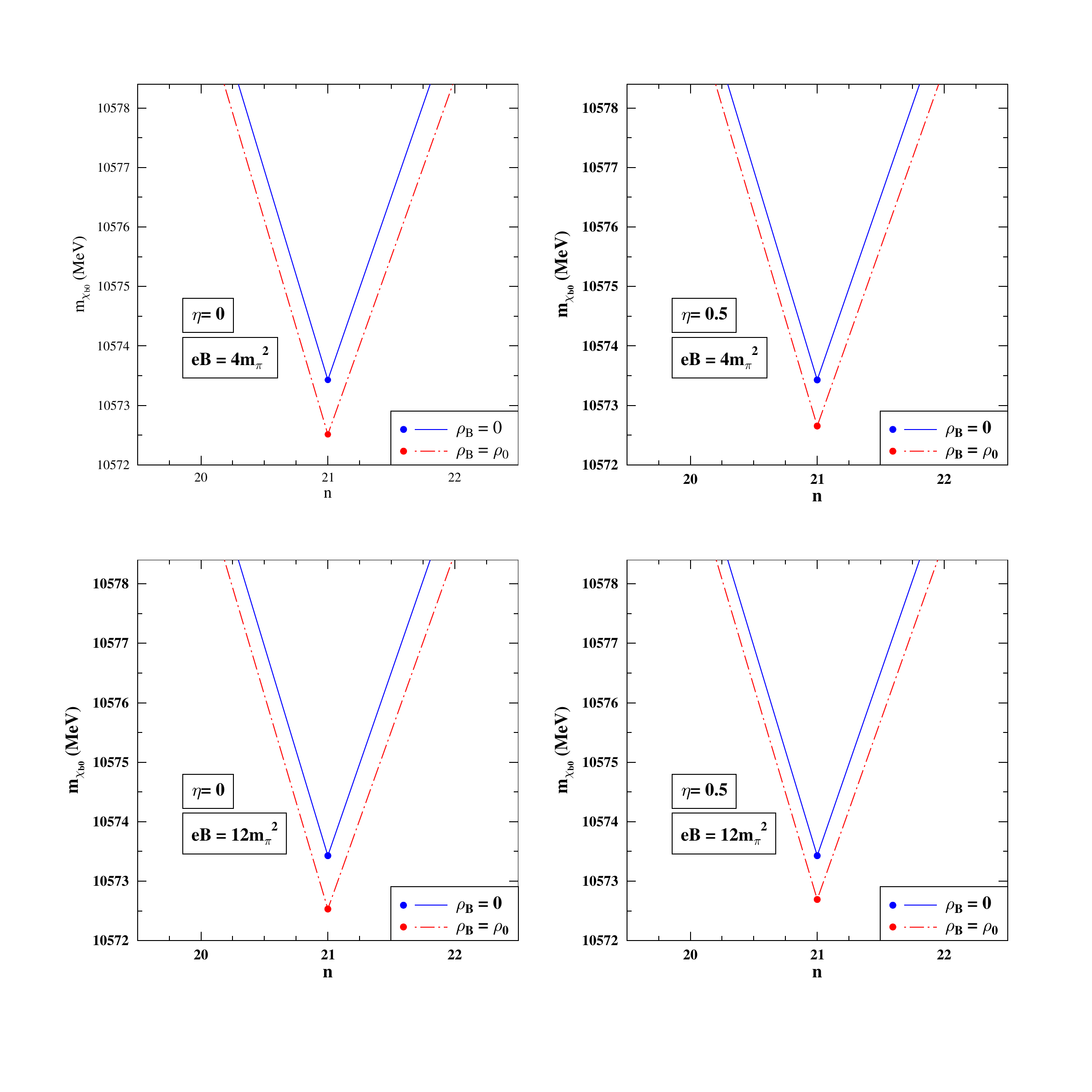}\hfill
\vskip -0.8in
    \caption{Masses (MeV) are plotted with variation in n, for the scalar 1P state, $\chi_{b0}$ at non-zero magnetic field (eB = 4$m_{\pi}^2$ and 12$m_{\pi}^2$), for $\rho_B = 0$ and $\rho_0$. Masses are plotted both at symmetric ($\eta$ = 0) and asymmetric ($\eta$ = 0.5) nuclear matter.}
    \label{fig:4a}
    \end{figure}
    \begin{figure}
    \includegraphics[width=1.1\textwidth]{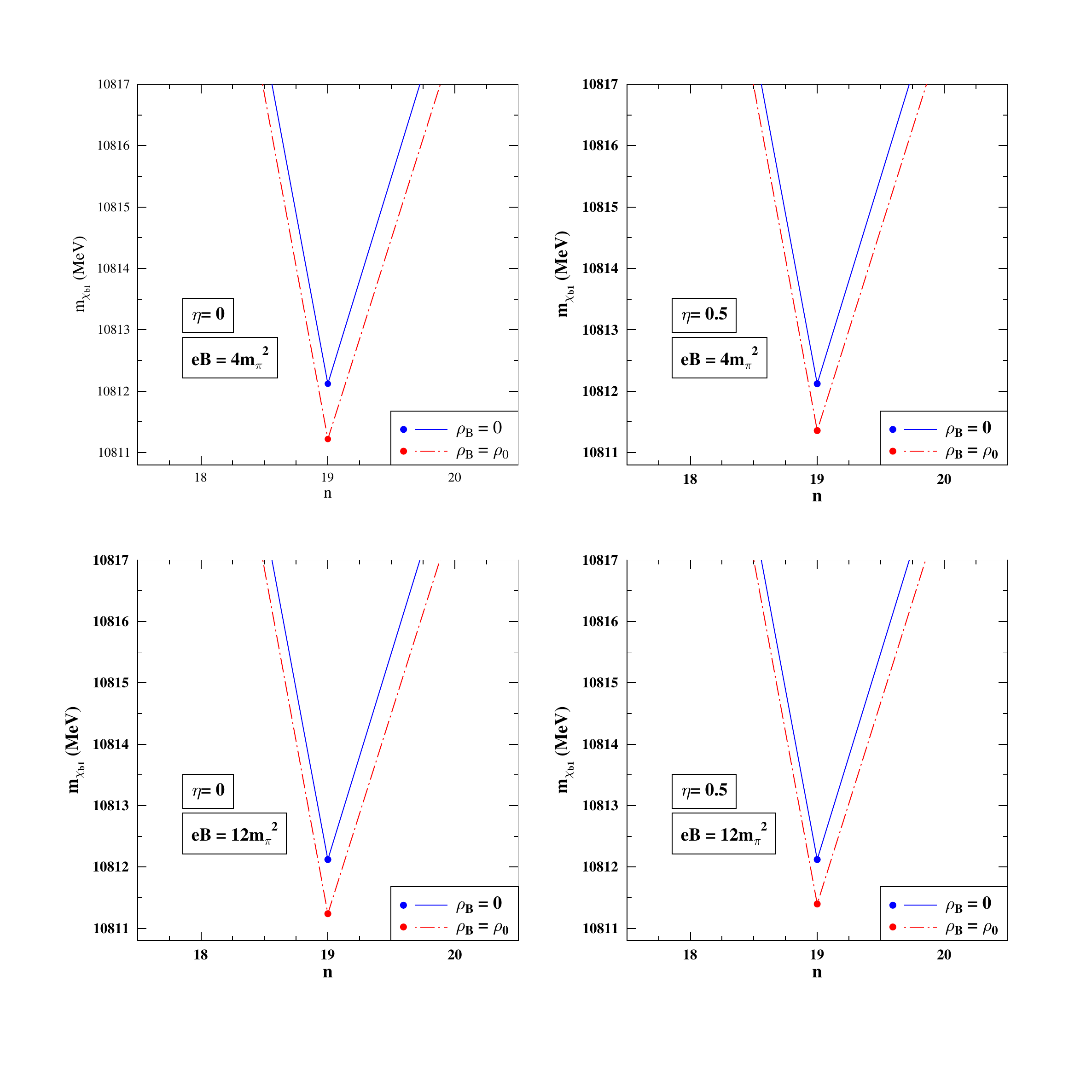}\hfill
\vskip -0.8in
    \caption{Masses (MeV) are plotted with variation in n, for the axial-vector 1P state, $\chi_{b1}$ at non-zero magnetic field (eB = 4$m_{\pi}^2$ and 12$m_{\pi}^2$), for $\rho_B = 0$ and $\rho_0$. Masses are plotted both at symmetric ($\eta$ = 0) and asymmetric ($\eta$ = 0.5) nuclear matter.}
    \label{fig:4b}
\end{figure}

\begin{figure}
    \centering
    \includegraphics[width=1.1\textwidth]{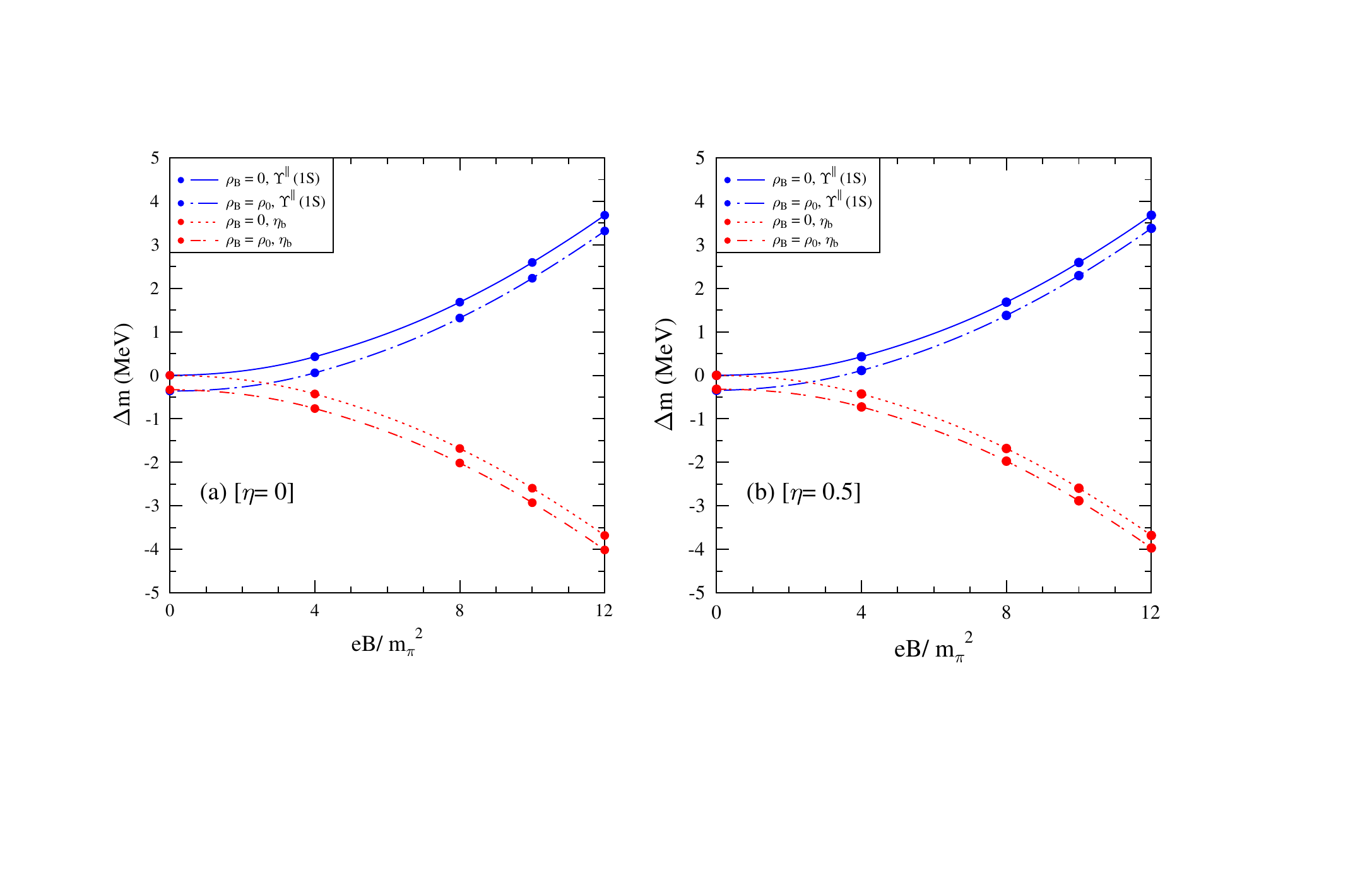}
    \hfill
\vskip -0.8in
    \caption{Mass shifts (MeV) are plotted as a function of magnetic field, eB (in units of $m_\pi^2$), for $\Upsilon^{||} (1S)$ and $\eta_b$ states by considering their spin-mixing effects in the presence of finite magnetic fields. Plot.(a) shows the $\eta = 0 $ case and plot.(b) is for $\eta = 0.5$. Mixing effects are considered at $\rho_B = 0$ and $\rho_0$. }
    \label{fig:5}
\end{figure}

\begin{table}
\begin{center}
\begin{tabular}{ |c|c|c|c| } 
\hline
Particle state & $\rho_B$ & $\eta=0$ & $\eta=0.5$ \\[2ex]
\hline \hline
 & $\rho_0$ & 9681.149 & 9681.16 \\ [2ex]
\cline{3-4}
& & \textbf{-0.33} & \textbf{-0.319}\\ [2ex]
\cline{2-4}
\textbf{$\eta_b$}& 2$\rho_0$ & 9680.762 & 9680.812 \\ [2ex]
\cline{3-4}
& & \textbf{-0.72} & \textbf{-0.67}\\[2ex]
\cline{2-4}
& 4$\rho_0$ & 9680.285 & 9680.396 \\[2ex]
\cline{3-4}
& & \textbf{-1.194} & \textbf{-1.083}\\ [2ex]
\hline \hline
 & $\rho_0$ & 9750.89 & 9750.903 \\[2ex]
 \cline{3-4}
 & & \textbf{-0.360} & \textbf{-0.346}\\ [2ex]
\cline{2-4}
\textbf{$\Upsilon (1S)$} & 2$\rho_0$ & 9750.317 & 9750.391 \\ [2ex]
\cline{3-4}
& & \textbf{-0.932} & \textbf{-0.857} \\ [2ex]
\cline{2-4}
& 4$\rho_0$ & 9749.574 & 9749.747 \\[2ex]
\cline{3-4}
&& \textbf{-1.675} & \textbf{-1.502} \\ [2ex]
\hline
\end{tabular}
\caption{Masses (MeV) and their shifts (MeV) from the vacuum values for lowest lying S-wave bottomonia at $\rho_B = \rho_0$, $2\rho_0$ and $4\rho_0 $; at zero magnetic field in the isospin symmetric and asymmetric nuclear matter.}
\label{table:2.a}
\end{center}
\end{table}

\begin{table}
\begin{center}
    \begin{tabular}{ |c|c|c|c| }
 \hline
 Particle state & $\rho_B$ & $\eta=0$ & $\eta=0.5$ \\[2ex]
\hline \hline
 & $\rho_0$ & 10572.541 & 10572.58 \\[2ex]
\cline{3-4}
&& \textbf{-0.89} & \textbf{-0.851} \\ [2ex]
\cline{2-4}
\textbf{$\chi_{b0} $} & 2$\rho_0$ & 10571.058 & 10571.252 \\ [2ex]
\cline{3-4}
&& \textbf{-2.370} & \textbf{-2.176} \\ [3ex]
\cline{2-4}
& 4$\rho_0$ & 10569.12 & 10569.571\\[2ex]
\cline{3-4}
&& \textbf{-4.310} & \textbf{-3.86} \\ [2ex]
\hline \hline
 & $\rho_0$ & 10811.248  & 10811.284 \\[2ex]
 \cline{3-4}
 && \textbf{-0.873522} & \textbf{-0.83756} \\ [2ex]
\cline{2-4}
\textbf{$\chi_{b1} $}& 2$\rho_0$ & 10809.788 & 10809.980 \\ [2ex]
\cline{3-4}
&& \textbf{-2.333} & \textbf{-2.142} \\ [2ex]
\cline{2-4}
& 4$\rho_0$ & 10807.878 & 10808.323 \\ [2ex]
\cline{3-4}
&& \textbf{-4.243} & \textbf{-3.798} \\ [2ex]
\hline
\end{tabular}
\caption{Masses (MeV) and their shifts (MeV) from the vacuum values for lowest lying P-wave bottomonia at $\rho_B = \rho_0$, $2\rho_0$ and $ 4\rho_0 $; at zero magnetic field in the isospin symmetric and asymmetric nuclear matter.}
\label{table:2.b}
\end{center}
\end{table}
Masses (MeV) are calculated at zero magnetic field in symmetric and asymmetric nuclear matter at the baryonic density $\rho_B$ = $\rho_0$, ${2\rho}_0$ and ${4\rho}_0$. They are listed in table.\ref{table:2.a} and table.\ref{table:2.b} for the S-wave and P-wave states respectively.
 In table.\ref{table:3}, the masses and their shifts in MeV, from the corresponding vacuum values are shown for the S-wave states [$\Upsilon (1S), \eta_b$], at eB = $4m_\pi^2$ and $12m_\pi^2$ both in symmetric ($\eta = 0.0$) and asymmetric ($\eta=0.5$) nuclear matter for $\rho_B=\rho_0$ and $2\rho_0$.
 Similar findings can also be made for the P-wave bottomonium ground states. Table.\ref{table:4}, illustrates the in-medium masses and their shifts from the vacuum, for the scalar $(\chi_{b0})$ and the axial vector $(\chi_{b1})$ states with the variation of $\rho_B$, eB, and $\eta$. 
 
 The mixing between the $1S$ $\Upsilon$ and $\eta_b$ states due to the spin-magnetic field interaction effect leads to a respective increase and decrease in their masses with increasing magnetic fields. The mass shifts for these states are shown in table.\ref{table:5} at $\rho_B=0, \rho_0$ and for $\eta =0, 0.5$ with the variation of magnetic fields, eB/$m_\pi^2$ = 4, 8, 10 and 12.

\begin{table}
\begin{center}
\begin{tabular}{ |c|c|c|c|c|c| } 
\hline
 \multirow{2}{4em}{eB} & \multirow{2}{4em}{$\rho_B$} & \multicolumn{2}{|c|}{\textbf{$\eta_b $}} & \multicolumn{2}{|c|}{\textbf{$\Upsilon (1S$)}} \\ [2ex]
 \cline{3-6}
  & & \textbf{$\eta=0$} & $\eta=0.5$ & $\eta=0$ & $\eta=0.5$\\ [2ex]
   \hline
 & \textbf{${\rho_0}$} & 9681.144 & 9681.181 & 9750.878 & 9750.932 \\[2ex]
\cline{2-6}
\textbf{$4m_\pi^2$} & $\Delta m$ & \textbf{-0.335} & \textbf{-0.298} & \textbf{-0.371} & \textbf{-0.317}\\[2ex]
\cline{2-6}
 & $ 2\rho_0 $ & 9680.774 & 9680.851 & 9750.337 & 9750.45\\[2ex]
\cline{2-6}
 & $\Delta m$ & \textbf{-0.705} & \textbf{-0.628} & \textbf{-0.911} & \textbf{-0.799}\\[2ex]
 \hline
 & $\rho_0$ & 9681.148 & 9681.193 & 9750.885 & 9750.948\\[2ex]
\cline{2-6}
 \textbf{$ 12m_\pi^2 $} & $\Delta m$ & \textbf{-0.332} & \textbf{-0.287} & \textbf{-0.364} & \textbf{-0.301}\\[2ex]
\cline{2-6}
 & $ 2\rho_0$ & 9680.776 & 9680.918 &	9750.338 & 9750.550\\[2ex]
\cline{2-6}
 & $\Delta m$ & \textbf{-0.703} & \textbf{-0.561} & \textbf{-0.911} & \textbf{-0.698}\\[2ex]
  \cline{2-6}
\hline
\end{tabular}
\end{center}
\caption{Masses (MeV) and their shifts (in MeV) for S-wave bottomonium ground states, at eB = $4m_\pi^2$ and $12m_\pi^2$ and for $\eta =$0 and 0.5.}
\label{table:3}
\end{table}
\begin{table}
\begin{center}
\begin{tabular}{ |c|c|c|c|c|c| } 
\hline
 \multirow{2}{4em}{eB} & \multirow{2}{4em}{$\rho_B$} & \multicolumn{2}{|c|}{\textbf{$\chi_{b0}$}} & \multicolumn{2}{|c|}{\textbf{$\chi_{b1}$}} \\ [2ex]
 \cline{3-6}
  & & \textbf{$\eta=0$} & $\eta=0.5$ & $\eta=0$ & $\eta=0.5$\\ [2ex]
   \hline
 & \textbf{${\rho_0}$} & 10572.513 & 10572.653 &10811.220&
	10811.359 \\[2ex]
\cline{2-6}
\textbf{$4m_\pi^2$} & $\Delta m$ & \textbf{-0.916} & \textbf{-0.775} & \textbf{-0.901} & \textbf{-0.763}\\[2ex]
\cline{2-6}
 & $ 2\rho_0 $ & 10571.113 & 10571.404 & 10809.842 & 10810.128\\[2ex]
\cline{2-6}
 & $\Delta m$ & \textbf{-2.316} & \textbf{-2.025} & \textbf{-2.280} & \textbf{-1.993}\\[2ex]
 \cline{2-6}
\hline
 & $\rho_0$ & 10572.531 &	10572.692 &10811.239 &	10811.397 \\[2ex]
\cline{2-6}
 \textbf{$ 12m_\pi^2 $} & $\Delta m$ & \textbf{ -0.897} & \textbf{-0.736} & \textbf{-0.883} & \textbf{-0.724}\\[2ex]
\cline{2-6}
 & $ 2\rho_0$ & 10571.114 &	10571.665 &	10809.843 & 10810.386\\[2ex]
\cline{2-6}
 & $\Delta m$ & \textbf{-2.315} & \textbf{-1.763} & \textbf{-2.279} & \textbf{-1.736}\\[2ex]
 \cline{2-6}
\hline
\end{tabular}
\caption{Masses (MeV) and their shifts (in MeV) for P-wave bottomonium ground states, at eB = $4m_\pi^2$ and $12m_\pi^2$ and for $\eta =$0 and 0.5.}
\label{table:4}
\end{center}
\end{table}

\begin{table}
\begin{center}
\begin{tabular}{ |c|c|c|c|c|c| } 
\hline
State & $\rho_B$ & \multicolumn{4}{|c|}{eB in units of $m_\pi^2$} \\ [2 ex]
\cline{3-6}
&& 4& 8  & 10 & 12 \\ [2ex]
\hline \hline
& 0 & 0.428 & 1.682 & 2.595 & 3.681 \\ [2ex]
\cline{2-6}
$\Upsilon (1S)$& $\rho_0 (\eta = 0)$ & 0.057 & 1.318 & 2.232 & 3.319 \\ [2ex]
\cline{2-6}
& $\rho_0 (\eta = 0.5)$ & 0.111 & 1.378 & 2.293 & 3.381 \\ [2ex]
\hline
& 0 & -0.428 & -1.682 & -2.595 & -3.681\\ [2ex]
\cline{2-6}
$\eta_b$ & $\rho_0 (\eta = 0)$  & -0.763 & -2.016 & -2.927 & -4.014 \\ [2ex]
\cline{2-6}
& $\rho_0 (\eta = 0.5)$ & -0.726 & -1.973 & -2.884 & -3.969\\ [2ex]
\hline

\end{tabular}
\caption{Mass-shifts (MeV) of 1S $\Upsilon$ and $\eta_b$ due to their mixing effects under spin-magnetic field interaction at vacuum ($\rho_B = 0$) and at $\rho_B = \rho_0$, for isospin symmetric ($\eta = 0$) and asymmetric ($\eta = 0.5$) matter. }
\label{table:5}
\end{center}
\end{table}

In QCD Sum Rule approach, masses of the lowest order resonances are calculated for a range of n-values (n : order of moments in QCDSR). The minimum point is considered to be the approximate physical mass of the corresponding resonance.The effects of the nuclear matter density, isospin asymmetry, and magnetic fields, on the bottomonia masses are shown in figures [\ref{fig:2a}-\ref{fig:4b}] with variation in n. As it is observed from these figures as well as from the mass shifts in tables [\ref{table:2.a}-\ref{table:4}], that among the other effects acting on the properties of bottomonia in nuclear medium, density has the significant contribution and that P-wave states have prominent mass drops than the S-wave states in the QCD sum rule approach calculations, under the same nuclear matter conditions. As it is shown in tables, [\ref{table:3}-\ref{table:4}], the mass shifts of  $\eta_b, \Upsilon(1S), \chi_{b0}, \chi_{b1}$ at nuclear matter saturation density, $\rho_B = \rho_0$ in isospin symmetric (extreme asymmetric) nuclear matter, $\eta=0 (0.5)$, for eB = 4$m_{\pi}^2$ are -0.335 (-0.298), -0.371 (-0.317), -0.916 (-0.775), -0.901 (-0.763), respectively, and at $\rho_B = 2\rho_0$ these are modified to -0.705 (-0.628), -0.911 (-0.799), -2.316 (-2.025), -2.280 (-1.993). \\
The dominant contribution from the magnetic fields, in mass shifts are obtained through the spin-mixing effects between the 1S states, $\Upsilon^{||}(1S)$ and $\eta_b$. This is due to the spin-magnetic field interaction at finite magnetic fields between the longitudinal component of the spin one and spin zero states of 1S bottomonia considered here. As can be seen from figure \ref{fig:5}, that the splitting between the $\Upsilon^{||}(1S)$ and $\eta_b$ are increasing with magnetic fields. In table \ref{table:5}, mass shifts for $\Upsilon^{||}(1S)$ and $\eta_b$, at $\rho_B = \rho_0$, for eB = 4$m_{\pi}^2$ (12$m_{\pi}^2$) are given for symmetric matter case, 0.057 (3.319) and -0.7634 (-4.014) respectively, and for extreme asymmetric case, 0.111 (3.381) and -0.726 (-3.969) respectively.     

\section{Summary}
To summarize the findings of the present work, the in-medium masses of both the 
S and P waves bottomonium states calculated using a QCD sum rule approach,
are observed to decrease with increasing baryonic density. The drop is 
more prominent for the P states. The effects of isospin asymmetry 
of the nuclear medium and the magnetic field, are seen to be appreciable 
at high densities. Both the asymmetry of nuclear medium and the strong magnetic fields effects coming via nucleons are seen to have negligible contribution 
in changing the in-medium properties as compared to the significant density 
contribution. The magnetic fields contribution is dominant while 
taking into consideration of the spin-magnetic field interaction 
between the 1S spin triplet and singlet states, which leads 
to an appreciable increase and decrease in the masses of the longitudinal 
component of $\Upsilon(1S)$ and  $\eta_b$. This might 
be observed as a quasi-peak at $m_{\eta_b}$ in the dilepton spectra
in non-central ultra-relativistic collisions at RHIC, LHC,
where the magnetic fields produced are huge.
For the zero magnetic fields, the bottomonium masses have dominant
effects from the baryon density, which can have consequences on the
production of the open and hidden bottom mesons in facilities
which probe the high density baryonic matter.

\end{document}